# Possible evidence from the flaring activity of Sgr A* for a starat a distance of ~3.3 Schwarzscild radii from the blackhole

Elia Leibowitz

School of Physics & Astronomy and Wise Observatory
Faculty of Exact Sciences
Tel Aviv University

eliamenl@gmail.com

Running title: **Modulation of Sgr A* flares by pacemakers**

**Abstract**

The frequent flaring events in the X-ray and the NIR radiation of Sgr A* seem not to be periodic in time. However, statistical regularities, here termed 'modulations by a pacemaker', are found in the recorded arrival times of both types of events. The characteristic time of the X-ray pacemaker is 149 min and that of the NIR pacemaker is 40 min. Their reality as derived from observed data can be accepted at larger than $4.6\sigma$ and $3.8\sigma$ levels of statistical confidence, respectively. These results can be interpreted as evidence for a star that revolves around the BH of Sgr A* in a slightly elliptical precessing orbit, at a distance of 3-3.5 Schwarzschild radii of the BH. The period of the X-ray pacemaker, which is not a periodicity of the flare occurrences themselves, is the epicyclic period of the star orbital motion. This is the time interval between 2 successive passages of the star through the pericenter of its orbit. The NIR pacemaker period is the mean sidereal binary period of the star revolution. The X-ray flares originate in episodes of intense mass loss from the star that occur preferably near the pericenter phase of the binary revolution. The NIR flares originate or are triggered by processes that are internal to the star. The radiation emitted in the direction of Earth is slightly modulated by the changing aspect ratio of the two components of the BH/star binary to the line of sight from Earth at the sidereal binary frequency.

# 1. Introduction

One of the characteristics of the Galactic center object Sgr A*, the supermassive blackhole (BH) and its close vicinity, is the occurrence of flares in its X-ray, as well as in its IR radiation (Baganoff et al. 2001; Genzel et al. 2003; Pouquet et al. 2008; Nielssen et al. 2013, hereafte N13; Ponti et al. 2015, hereafter P15). These are short outbursts of otherwise weak emission, lasting a few to a few tens of minutes during which the radiation intensity increases by factor of up to a few hundreds of its quiescence value (Dodds-Eden et al.2011; Witzel et al. 2012; Barriere et al. 2014; Zhang et al. 2017; Haggard et al. 2019). The average rate of these events is about 1 or 1.1 per day for the X-ray (Markoff 2010; Li et al. 2015; N13). The NIR flares seem to occur about 4-5 times a day (Witzel et al. 2012; Hora et al. 2014). Do et al. (2019) have reported on a recent significant change in this pattern of the NIR radiation. A brief comment regarding this observation is made in Section 8.3.2.

A few mechanisms have been proposed as models for the physical process or processes that generate the radiation outbursts of the flares (Trap et al. 2011; Zhang et al. 2017; Ripperda et al. 2020 and references therein). However, the mean rates of occurrence of either the X-ray or the NIR flares are not yet well understood.

In two earlier papers (Leibowitz 2017, 2018, hereafter L1,L2) I reported on a statistical regularity, which I call 'modulation by a pacemaker', that can be identified in the distribution along the time axis of mid points of 71 X-ray flares of the system. I found that X-ray flares tend to be detected close to points on the time axis that belong to a grid of constant interval of $P_X = 0.1032\ days$ between any two of its neighboring points. In the following discussions, unless indicated otherwise, all time intervals will be expressed in units of days and frequencies in $1\ day^{-1}$ units. In these earlier papers I also suggested that these regularities may be evidence for a star that revolves around the BH and serves as the pacemaker.

In this paper I present an improved analysis of the time series of the mid points of a set of X-ray flares that is slightly larger than the sets analyzed in the previous two papers. Also, the search for a pacemaker signal and the tests of its statistical significance are here performed over a frequency search interval that is 2.5 times larger than in L2. I also report here for the first time on the results of a similar analysis performed on a large set of NIR flares of the object that were recorded during the years 2000-2017. The analysis presented here reconfirms the findings in the previous papers regarding the X-ray flares. It also reveals that the times of the NIR flares are also modulated statistically by a pacemaking process, much like the X-ray flares. The cycle of the NIR pacemaker is, however, $P_{IR} = \sim 0.028$, clearly different from that of the pacemaker of the X-ray flares. This new result is very much consistent with the model suggested in the previous 2 papers as interpretation of the regularities in the X-ray events.

The outline of this paper is as follows. Section 2 presents briefly the Frequency Dispersion Diagram (FDD), the basic tool of my statistical analysis, the detailed of which were introduced in L1 and L2. It also presents 3 statistical tests of the confidence level at which the results can be accepted. In Section 3 I present the X-ray data that are added here to the list of the X-ray flares analyzed in L2. The results of the FDD analysis on the extended X-ray data are presented in section 4, as well as the confidence level that can be ascribed to them. In Section 5 I present the NIR data analyzed in this work and Section 6 presents the results of the statistical analysis of them. Section 7 contains some discussion of the relation between the statistical analysis that is presented in this work and some of other statistical analyses of the flaring activity of the object that have been performed and published in the literature. Section 8 presents the model that I propose for interpreting the finding in the flares timing. The basic equations that underlie the model are given, as well as the value of the model parameters implied by applying them on the observational data. Section 9 contains some critical and concluding remarks.

# 2. Revealing a Pacemaker in a Time Series and Its Statistical Significance

## 2.1. The Frequency Dispersion Diagram

Computation of a Frequency Dispersion Diagram (FDD) is a numerical procedure to search in a given series TN of N points on the time axis the statistical regularity which I term modulation by a pacemaker. The method is described in details and the mathematics involved is presented in section 3 of L2.

The FDD is defined on a sample of uniformly distributed frequencies covering a given frequency search interval I. The frequency $F_{min}$ of the minimum point in the FDD plot is the frequency of the grid on the time axis with respect to which the N time points are grouped most tightly among the grids of all the frequencies in the I search intervals. The StD of the N time points around the tick marks of the $F_{min}$ grid is referred to as the S parameter of the given set. The number $F_{min}$ and its reciprocal number $P_{min} = \frac{1}{F_{min}}$ are the frequency and the period of the pacemaker. Note, however, that they are not the frequency or a periodicity in the distribution on the time axis of the set TN time points themselves.

In order for $F_{min}$ to be of any statistically meaningful significance one must show that there is only small False Negative Probability (FNP) that the regularity exposed by the FDD procedure is not a matter of chance coincidence. The smaller FNP is, the higher is the statistical confidence that can be given to the rejection of the null hypothesis that the N numbers of the TN set are in fact random, in a sense explained in the following section. This is done by way of simulations.

## 2.2. Tests of Statistical Significance

### 2.2.1. Simulating a Flare by Randomizing the Hour of Its Occurrence.

As already mentioned in Section 1, the average rate of X-ray flares of Sgr A* is about 1.1 per day. Here we are interested in the statistics of the times of events on the scale of one day to half an hour. I therefore create sets of N pseudo-observed (PO) events by considering the time UT 00:00 of each of the days of the observed events, adding to it a number selected randomly from a square probability density distribution over the [0,1] interval, using the RAND command of the MATLAB environment. This random number replaces the hour of the day when the observed event has actually occurred. In this way the statistics of the occurrences of the recorded real flares on time scale longer than 1 day is preserved. A set of randomized, pseudo-observed events so created is here referred to as an SN set.

I also simulated the observed set with numbers that are random in a more radical way. They are chosen randomly from a rectangle probability density distribution over the entire historic time interval, or over the subsections of it along which X-ray observations have been taking place. The estimated FNP values based on these simulations are smaller than the ones reported about here.

### 2.2.2. S-test

A straight forward estimate of FNP in rejecting the null hypothesis can be obtained by computing the FDD function of a large number $N_S$ of SN sets of PO events. The required estimate is $FNP \approx \frac{K_S}{N_S}$, where $K_S$ is the number of SN sets for which the S value is equal or smaller than the S value of the set of the real data.

### 2.2.3. G-test

We shall see in the following sections that the sets of the observed times of flares possess another statistical property, not shared by the PO sets. On the frequency axis, within a small neighborhood w around the deepest minimum in the FDD, there is a particularly large number k of frequencies of other minima, all among the group of the K deepest ones in the FDD. This feature is a mark of stability or persistence of the pacemaker signal in the recorded time series. The FNP in rejecting the null hypothesis with respect to this property can be estimated on the basis of the fraction of the sample of simulated sets, in the FDD of which within the±w neighborhood of the deepest minimum there is a number k of other minima that is larger or equal to the corresponding one found in the FDD of the real data set, with the same w and K values.

A plot of the k value vs. the S values found in FDDs of a large number of SN sets reveals that there is a linear correlation between these 2 parameters, manifested in the plot as a linear regression line with a negative slope. Detrending the distribution of the k values of the real data set and of the sample sets by removing the regression line between these 2 parameters, allows us to regard the G-test as independent of the S-test.

### 2.2.4. C-test

The determination of the time of each recorded flare depends solely on information related to the very same flare, its beginning and end times for the X-ray set or the time of its peak count rate for the NIR flares (see Section 5.1). It is entirely independent of the determination of the time of any other flare. Therefore any subset TNa of TN is independent of any other, mutually exclusive subset TNb of TN.

In the FDD of TNa computed on a dense equidistant grid of frequencies covering the interval I on the frequency axis, let $f_a$ be the frequency of the minimum in the FDD that has the ordinal number $n_a$ in the list of the minima ordered by their depth. Let $f_b$ and $n_b$ be the corresponding numbers in the FDD of TNb computed on the same frequency grid. The Cartesian product of the frequencies of the $n_a$ deepest minima in the FDD of TNa and the frequencies of the $n_b$ deepest minima in the FDD of TNb contains $n_a \times n_b$ pairs of numbers.

Under the null hypothesis that the 2 sets TNa and TNb are independent of one another, for a given number $\delta \leq \frac{I}{2}$, the probability that for at least one pair the absolute difference between the values of the pair members will be $\leq \delta$, is

(1) $$Pr = 1 - \left(1 - \frac{2\delta}{I}\right)^{[n_a \times n_b]}.$$

If $f_a$ and $f_b$ are within the uncertainty intervals in the values of one another, this expression may serve as an estimate of FNP of rejecting the null hypothesis and accepting that the set TN is modulated by a pacemaker of a frequency within that uncertainty interval.

C-test is independent of S-test since it is concerned with the distribution along the frequency axis of frequencies of deep minima in the FDD while S is concerned with the depth of the deepest minimum in the FDD.

## 3. X-ray Data

Table 1 in L2 presents a list of times of mid points of 71 X-ray flares analyzed in that paper. Table 1 of this paper lists the times of midpoints of additional 10 X-ray flares that I found in the literature since publication of L2. Whenever the data source of a flare does not provide its beginning and end times, these were determined by eye from the published light curve (LC) of the event.

| | | | | |
|---|---|---|---|---|
| 4194.7408* | 6123.1583* | 6144.3192* | 6727.2455# | 6750.2104# |
| 7157.0253% | 7582.4560@ | 7588.1367@ | 7950.4852@ | 7951.0811@ |

Table 1: Times in HJD-2450000 of midpoints of 10 X-ray flares of Sgr A*. Times of another 71 flares are presented in Table 1 of L2. The symbols indicate sources of data: (*) -  Ponti et al. (2015); (#) - Mossoux et al. (2016); (%) – Fazio et al. (2018); (@) – Boyce et al. (2019).

The list of the 71+10 X-ray times is here referred to as the TX81 set.

## 4. Analysis of Times of X-ray Flares

### 4.1 FDD of TX81

The search for an effect of a pacemaker on the distribution of the TX81 recorded times of the 81 X-ray flares is here conducted within a frequency search interval [0.8333-50], as compared to the much narrower interval [2-20] considered in L2. The corresponding period search interval is [0.02-1.2]. The upper period limit is determined by the known statistical fact that the mean occurrence frequency of the X-ray flares of Sgr A* is $1.1\,day^{-1}$ (N13, P15, Fazio et al. 2018). The lower limit is about half the time interval between the two closest flares in TX81.

The number of frequencies that I consider within the search interval $I_f$ is $n_f = 1750000$, so that $\frac{I_f}{n_f} \cong \frac{1}{6}\left(\frac{1}{L}\right)$, where $\left(\frac{1}{L}\right)$ is the formal spectral resolution of a time series of length L. Here $L \approx 6000\,day$. This choice of $n_f$ insures that the scan over the frequency search interval will not leave undetected any significant frequency within the interval. This equation will also be used in the following sections for the determination of the $n_f$ value in the computations of FDDs of time series of different length L.

The application of FDD on TX81 yields a pacemaker with virtually the same frequency as found in L2 with TX71, namely, $F_X = 9.6897$, corresponding to the period $P_X = 0.1032 = 148.6\,minutes$. The dispersion of the TX81 points around tick marks of the $F_X$ grid is $S_X = 0.1976$. With the bootstrap method (Efron and Tibshirani, 1993) I estimate an uncertainty of $\pm 0.03$ in the value of $F_X$.

### 4.2. Significance of the $F_X$ Frequency

#### 4.2.1 Qualitative

In L2 it was found that when the flare times analyzed there are expresses in HJD units the claim of modulation of the data by a pacemaker, as revealed by the FDD analysis, can be accepted with a level of statistical confidence that is higher than when times are expressed, as in most sources of the data, in JD or UT units. The same is true also for the extended set of 81 events discussed in this paper. This may be regarded as some qualitative evidence that the preferred grouping of the 81 X-ray flare times around the tick marks of the $F_X$ grid is likely to be an expression of some external reality and not a sheer coincidence.

### 4.2.2. S-test and G-test

Applying S-test on a sample of 1002 SX81 sets of PO events I find 17 sets with $S \leq S_X = 0.1976$. The implied FNP is $FNP_X(Stest) \approx \frac{17}{1000} = \frac{1}{59}$. G-test applied on this sample with K=500 and $0.014 \leq w \leq 0.03$ reveals no SX81 set with $k \geq k_X$ where $k_X$ is the k parameter, introduced in section 3.2.3, that is found for the real data. This sets an upper limit of $\frac{1}{1000}$ for the value of the FNP based on this test.

As these 2 tests may be regarded as independent of each other, we obtain the combined estimate $FNP_X^{(1)} \leq \left(\frac{1}{59}\right)\left(\frac{1}{1000}\right) = \frac{1}{59000}$ of the probability of falsely rejecting the null hypothesis. The claim that the midpoint times of the 81 X-ray flares of Sgr A* are modulated by a pacemaker with the period $P_X = 0.1032\ days$ may therefore be accepted at a $4.3\sigma$ level of confidence.

### 4.2.3. C-test

I divide the set TX81 into 3 subsections. TX24 is the subset consisting of the first 24 time points of TX81, recorded between 2000 October and 2011 April. TX41 consists of the next 41 TX81 events, recorded between 2012 February and October. These are mostly the fruits of the Chandra X-ray Observatory's 2012 Sgr A∗ X-ray Visionary Project (N13). TX16 is the last 16 events of TX81 recorded between 2013 August and 2017 July (see Figure 1 in L2). I apply the FDD analysis on these 3 subsets with the following results: The frequency $f_a$=9.7068 is found at the $n_a$=165 position in the list of the deepest minima in the FDD of TX24. For TX41 the deepest minimum, $n_b$=1, is at the frequency $f_b$=9.6778 with S=0.1680. For TX16, the frequency $f_c$=9.6955 is found as the 75th deepest minimum in the FDD, $n_c$=75.

Applying S-test on the TX41 subset and on a sample of 500 SX41 sets we can estimate for this set $FNP(Stest) = \frac{4}{500}$. Application of G-test with the parameters values $0.013 \leq w \leq 0.03$ and K=12 on TX41 and on the same sample of simulated sets provides an independent estimate $FNP(Gtest) = \frac{21}{500}$ . The probability that the finding of the $f_b$ pacemaker in the TX41 data set is a random occurrence is the product of these 2 probabilities, $FNP_b = \frac{1}{2976}$. Using equation (1), with the parameters $n_a$, $f_a$, $n_b$, $f_b$ and I=49.167 we can estimate the probability of a random coincidence, within $\pm w = \pm 0.029$, of $f_a$ with $f_b$ and of $f_c$ with $f_b$ as $FNP_a = \frac{1}{5.65}$ and $FNP_c = \frac{1}{19}$ , respectively. The overall probability of falsely rejecting the null hypothesis and accepting that TX81 is modulated by a pacemaker of the frequency $F_X$ can therefore be estimated as $FNP_X^{(2)} \leq FNP_a \times FNP_b \times FNP_c = \frac{1}{319400}$ , corresponding to a statistically confidence level of $4.66\sigma$.

The product of $FNP_X^{(1)}$ and $FNP_X^{(2)}$ cannot be utilize as further improvement on the overall FNP estimation. $FNP_X^{(2)}$ derived in this subsection is not independent of $FNP_X^{(1)}(Stest)$ of the previous one since TX41 is a subsection of TX81. However, the 2 estimates made in section 4.2.2 and 4.2.3 are clearly not statistically equivalent to one another.

In this work we are considering 14% more X-ray events than in L2. More importantly, here the frequency search interval is 2.5 times wider than in L2. It is reassuring to find that the pacemaker frequency, as well as the statistical confidence level that can be attributed to it as derived here, are similar to those reported about in the previous L2 paper.

## 5. NIR flares

A natural step following the discovery of the statistical regularity in the timing of the X-ray flares of Sgr A* is an attempt to find out whether or not a similar signal can be revealed in the timing of the IR flares of the system. For this purpose I collected from the literature all the reports that I could find on observed IR flares that provide also information on the exact time of the events.

## 5.1. Spitzer Space Telescope Data

One major source of NIR flare data is in the paper of Witzel et al. (2018). These authors present eight 24-hour epochs of continuous monitoring of Sgr A* performed by the *Spitzer* space telescope between the years 2013 and 2017. Four of them, binned into 100 points per bin are shown in figure 1 as the solid curve, with HJD-2450000 as the x-axis variable. The y axis expresses flux density in units of mJy, as in the original paper.

Unlike the case of the X-ray flares, the times of the IR events are taken here as the x coordinates of the apparent peaks of the curve above the y=0 line (see Section 8). In this way I found 149 events, the times of which, HJD-2450000, are presented in Table 2. They are marked by vertical lines in Figure 1 and referred to as the TS149 set. To each event I associate the parameter $h$ which is the height of the peak above the y=0 line.

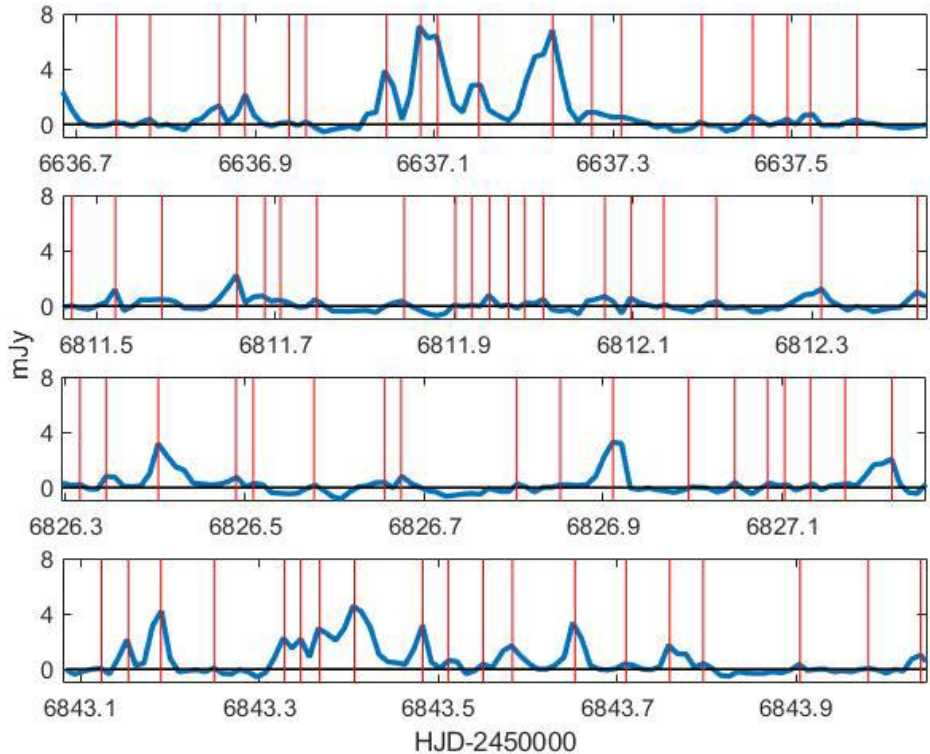

Figure 1: Solid curve in all panels is a NIR light curve of Sgr A* as recorded in four 24 hr epochs of continuous monitoring of the object at 4.5 $\mu m$ by the *Spitzer* space telescope between the years 2013 and 2017 (Witzel et al. 2018). Vertical lines mark the peaks of 75 features identified here as flares. Another 74 peaks in the additional four LCs recorded by *Spitzer* are not presented graphically in this paper.

| | | | | | |
|---|---|---|---|---|---|
| 6636.7447 | 6636.7824 | 6636.8598 | 6636.8889 | 6636.9373 | 6636.9566 |
| 6637.0466 | 6637.0844 | 6637.1037 | 6637.1492 | 6637.2324 | 6637.276 |
| 6637.3089 | 6637.3989 | 6637.456 | 6637.4947 | 6637.5198 | 6637.5721 |
| %6811.4721 | 6811.5205 | 6811.5727 | 6811.6569 | 6811.6879 | 6811.7053 |
| 6811.746 | 6811.8437 | %6811.9008 | %6811.9192 | 6811.9395 | 6811.9608 |
| 6811.9792 | 6811.9995 | 6812.0682 | 6812.0973 | 6812.134 | 6812.1931 |
| 6812.3102 | 6812.4176 | 6826.3156 | 6826.3447 | 6826.4027 | 6826.4898 |
| 6826.5092 | 6826.5769 | 6826.6563 | 6826.6747 | 6826.8034 | 6826.8527 |
| 6826.9118 | %6826.996 | 6827.0473 | 6827.084 | 6827.1034 | 6827.1324 |
| 6827.1711 | 6827.2234 | %6843.1234 | 6843.1534 | 6843.1902 | 6843.2492 |
| 6843.3276 | 6843.346 | 6843.3673 | 6843.406 | 6843.4824 | 6843.5115 |
| 6843.5502 | 6843.5831 | 6843.6527 | 6843.7098 | 6843.7582 | 6843.796 |
| 6843.9044 | %6843.9808 | 6844.0389 | 7582.3224 | 7582.3427 | 7582.3718 |
| 7582.4676 | 7582.5073 | 7582.5663 | 7582.5847 | 7582.6147 | 7582.6331 |
| 7582.6631 | 7582.7415 | 7582.7705 | 7582.8015 | 7582.8498 | %7582.9089 |
| 7582.9379 | %7582.9863 | 7583.025 | 7583.0831 | 7583.1905 | 7588.0305 |
| 7588.0885 | 7588.1369 | 7588.164 | 7588.2627 | 7588.2831 | 7588.3208 |
| 7588.3595 | 7588.4166 | 7588.436 | 7588.4553 | 7588.5269 | 7588.5869 |
| 7588.646 | 7588.675 | 7588.7127 | 7588.7408 | 7588.9266 | 7588.9663 |
| 7950.4866 | 7950.5147 | 7950.5718 | 7950.7198 | 7950.7479 | 7950.7973 |
| 7950.8447 | 7950.8834 | 7950.9134 | %7950.9502 | 7951.014 | 7951.0895 |
| 7951.1979 | 7951.2453 | 7951.2753 | 7951.3131 | 7951.3334 | 7951.4418 |
| 7960.4856 | 7960.6124 | 7960.6695 | 7960.7189 | 7960.7556 | 7960.7769 |
| 7960.894 | 7961.0498 | 7961.0798 | 7961.1098 | 7961.1495 | 7961.1679 |
| 7961.2076 | 7961.255 | 7961.3334 | 7961.3895 | 7961.4485 | |

Table 2: Set TS149 - times (HJD-2450000) of peaks of 149 NIR flares of Sgr A* recorder by the *Spitzer* space telescope between the years 2013-2017 (Witzel et al. 2018). The % symbol marks the 9 weakest flares (see Section 6.2.3).

## 5.2. Ground-based Data

A second source of data on timing of NIR flares is a collection of records of observations performed at a number of ground-based observatories, between the years 2002 and 2015, that I was able to gather from the literature. I refer to them as G flares. Unlike the data of *Spitzer* and of the X-ray flares, the published information regarding G flares is nearly always in the form of a graphical plot of a LC, and the UT or JD time of the beginning of the observing run along which the depicted LC was recorded. It was left to the onlooker to determine by eye the point of maximum of each feature in the plot that is considered a flare. The lack of digital data on most of the G flares makes this selection process a subjective task to some extent. It seems, however, that the difference between the time of peaks selected by different people would hardly be larger than a very few minutes. But even where there is some unknown personal bias in the selection by an individual person, it is next to impossible that it introduces into the numbers any statistical regularity in the time intervals between flares. The determination of the time of the peak of each flare was based solely on the plot of that particular flare, entirely independent of information regarding any other flare.

The reference zero intensity level of each flare had also to be determined by eye estimate in order to assign to each flare a value representing its height h in some arbitrary units. The h numbers suffer even more severely from the personal subjective biases in the selection process since the resolution in the time and intensity coordinates in the published graphical plots are far poorer than in the digital LC of *Spitzer*. However, while these limitations introduce some uncertainty in the precise ordering of flares that have comparable height, the h parameter is still adequate to differentiate between high and low peaks.

The list of the times HJD-2450000, of the peaks of 120 G flares, referred to as TG120, is presented in Table 3.

| | | | | | |
|---|---|---|---|---|---|
| 2516.5534[1] | 2768.8233[1] | 2805.6517[1] | 2806.7697[1] | 2810.5044[2] | 3169.8697[14] |
| 3192.9306[9] | 3193.4778[7] | 3193.5701[7] | 3193.6653[7] | 3194.6507[3] | 3248.9428[4] |
| 3250.6947[4] | 3250.7433[4] | 3250.9628[4] | 3252.6793[4] | 3252.8175[4] | 3503.8486[7] |
| 3504.8514[7] | 3506.7536[7] | 3506.8175[7] | 3541.7413[7] | 3576.6568[7] | 3579.79[8] |
| 3579.8657[8] | 3581.5607[7] | 3581.6267[6] | 3582.7447[7] | 3859.0174[8] | 3859.0535[8] |
| 3887.7868[14] | 3906.91[8] | 3907.8989[8] | 3907.9468[8] | 3915.7252[7] | 3933.7899[8] |
| 4001.5122[7] | 4002.517[7] | 4179.9141[7] | 4191.8125[10] | 4192.0431[10] | 4192.1195[10] |
| 4192.3779[10] | 4193.137[10] | 4193.2529[10] | 4193.3884[10] | 4193.7315[10] | 4193.8301[7] |
| 4193.8988[7] | 4194.1169[10] | 4194.7434[10] | 4195.1163[10] | 4195.2045[10] | 4195.9338[10] |
| 4195.9921[10] | 4196.3303[10] | %4196.7311[10] | 4196.9866[10] | 4197.1248[10] | 4198.1215[10] |
| 4198.177[10] | 4235.816[6] | 4237.8175[7] | 4238.987[8] | 4239.7398[7] | 4299.5427[13] |
| %4299.6024[13] | 4300.4628[13] | 4300.5704[13] | %4300.6995[13] | 4301.469[13] | 4301.5523[13] |
| %4301.6787[13] | 4303.5258[13] | 4303.673[13] | 4304.5278[13] | 4304.7313[13] | 4304.7702[13] |
| 4304.807[13] | 4305.7292[7] | 4538.8705[7] | 4597.8248[7] | 4611.7998[14] | 4611.913[14] |
| 4612.7943[11] | 4613.7763[14] | 4613.8367[14] | 4616.8562[14] | 4618.8834[14] | 4620.7314[5] |
| 4620.782[5] | 4620.839[5] | 4620.8946[5] | 4633.6031[7] | 4683.561[7] | 4724.5407[7] |
| 4725.5746[7] | 4921.9042[7] | 4922.8126[12] | 4924.8746[12] | 4969.9072[14] | 5015.6675[7] |
| 5015.7932[7] | 5016.7737[7] | 5017.5716[7] | 5017.7841[7] | 5018.841[7] | 5055.6147[7] |
| 5094.4819[7] | 5095.527[7] | 5284.863[7] | 5708.7068[14] | 5708.7901[14] | 6064.7272[14] |
| 6727.216[15] | 6727.4007[15] | 6750.2238[15] | %6750.8322[15] | 6751.8767[15] | 7157.0448[16] |

Table 3: Set TG120, the HJD-2450000 times of 120 peaks of IR flares of Sgr A* recorded by Ground-base observations between the years 2002 and 2015. Numbers in brackets refer to data sources as follows:
[1] Genzel et al. (2003); [2] Eckart et al. (2004); [3] Eckart et al. (2006); [4] Yusef-Zadeh et al. (2006); [5] Eckart et al. (2008b); [6] Eckart et al. (2008a); [7] Witzel et al. (2012); [8] Do et al. (2009); [9] Meyer et al. (2009); [10] Yusef-Zadeh. et al. (2009); [11] Kunneriath et al. (2010); [12] Trap et al. (2011); [13] Haubois et al. (2012); [14] Shahzamanian et al. (2015); [15] Mossoux et al. (2016); [16] Fazio et al. (2016). The % symbol marks the 5 weakest flares (see Section 6.2.3).

## 5.3. Combined set

The combination of the 2 sets TS149 and TG120 makes the set TIR269, a list of the times of peaks of all the 269 IR flares of Sgr A* analyzed in this work.

## 6. Analysis of the NIR Data

### 6.1. The $F_{IR}$ Frequency

Figure 2 is the FDD of the TIR269 set computed in the frequency range and sampling rate as done in Section 4.1 for the X-ray flares. The most outstanding feature in the figure is the deep minimum around the frequency $f \cong 1$. This feature is contributed to the FDD of TIR269 by its subset TG120, the list of flare times recorded by the ground-based observations. It is even more pronounced in the FDD of the TG120 component of TIR269, and it is entirely missing from the FDD of TS149, the other component.

The $f \cong 1$ feature is none other than the signal of the operation of the most powerful pacemaker, namely, Earth rotation, that modulates detections by all optical and IR observations performed from a single or neighboring ground observing sites. It is gratifying to see that our pacemaker search routine has easily "discovered" this universal pacemaker. The observations by the three space telescopes *Chandra*, $XMM-Newton$ and *Spitzer* are of course unaffected by the diurnal day/night variations.

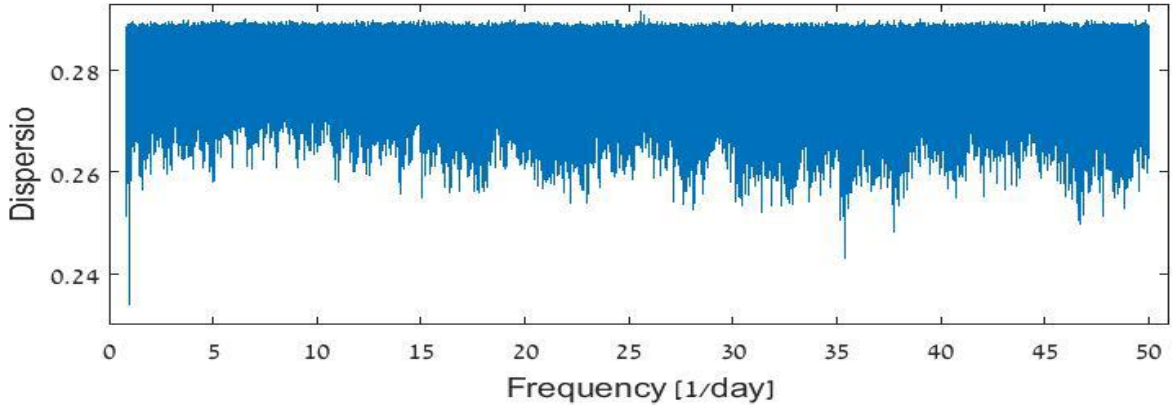

Figure 2: FDD of TIR269, the times of the peaks of 269 NIR flares of Sgr A* observed between the years 2002 and 2017.

In the frequency region f>1.2 the recorded flare timings are not affected directly anymore by the Earth rotation pacemaker. In further discussions, unless stated otherwise, the FDDs that I shall refer to are computed in the period interval [0.02-0.8333], (frequency interval [1.2-50]) sampled by 1700000 equidistant frequencies.

The most prominent feature to the right of the frequency $f \approx 1$ in Figure 2 is the minimum at $f_1 = 35.3914$ with the dispersion parameter value S=0.2449. The second-deepest minimum is at $f_2 = 35.3941$ with S= 0.2492 and the 4th one is at $f_3 = 35.4005$ with S= 0.2499. As done in Section 4.1, by the Bootstrap method an uncertainty of ±0.5 can be estimated in these f values. They can therefore be considered as representing one and the same frequency, for which I take as a representative number the value $F_{IR} = 35.395$. The differences are due to the intrinsic non-coherent nature of the pacemaker itself. This will be further discussed in Section 8.

## 6.2. Significance of the $F_{IR}$ Frequency

### 6.2.1. Qualitative

As noted in Section 4.2.1 regarding the X-ray flares, in the FDD of the IR flares too, when expressing the peak times in JD units rather than HJD, the deepest minimum is found at the same frequency $F_{IR}$, but the dispersion parameter is larger than the one obtained with HJD times. Also the statistical significance of the $F_{IR}$ frequency, as will be established in the following discussion, is much reduced when JD times are employed. This may be regarded as a qualitative evidence for some connection of $F_{IR}$ with external reality.

As already noted in Section 5.1, unlike the case of the X-ray flares, where the times of the flares midpoints were analyzed, here I consider the times of the peaks of the IR flares as the events to be analyzed. Applying the FDD process on midpoints of the IR flares yields the same $F_{IR}$ frequency but at a much lower statistical significance level. This point will be further discussed in Section 8.

### 6.2.2. S-test

I apply S-test, described in Section 3.2.2 on $F_{IR}$, the frequency of the deepest minimum in the FDD of TIR269. In the FDDs of a sample of 150 SIR269 sets of simulated events I find 28 with a minimum that is deeper than the one found in the FDD of the real data. This allows us to estimate the False Negative Probability that TIR269 is not a member of the simulated SIR269 group as $FNP(Stest) \approx \frac{1}{5.4}$.

In the simulation of the TG120 set, the subset of TIR269 that consists of the flares recorded in ground-based observations, I also took one additional step. As discussed in Section 6.1, all events of TG120 are confined, by the pacemaker Earth rotation, to only about half a day, centered around the time of the meridian crossing of the object in the sky of the observer. For any event t(j) on day d(j) in the observed set TG120, the meridian crossing time was taken as the time of the tick mark m(j) of the grid of the frequency f=1.0027, the frequency of the deepest minimum in Figure 2, that is the nearest one to t(j). In this additional, second type of simulation, the time of the observed event t(j) is replaced not by UT 00:00 of d(j) plus a random number from the [0,1] interval, but by the date m(j) plus a random number from a rectangular distribution over the [-0.25,0.25] interval. It is hardly in need to say that in the FDDs of such simulated subsets SG120, when computed over the period interval [0.02-1.2] the $f \approx 1$ minimum is the most pronounced feature, as it is for the real data TG120. The statistics based on sets of PO events with this more restrictive simulation is not significantly different from those reported on here.

### 6.2.3. C-test

C-test was presented in Section 3.2.4. Here I consider the two independent lists of events TS149 and TG120. In the FDD of TS149 the frequency $f_a = 35.3967$ is that of the $n_a = 7$ minimum in the list of the deepest ones in the FDD of this set. In the FDD of the TG120 set, a minimum at $f_b = 35.4006$ is found in the $n_b = 17$ position. Using equation (1) we can estimate the probability $NFP(Ctest) \approx \frac{1}{105}$.

Combining the results of the 2 independent tests S-test and C-test, we get $FNP \approx \frac{1}{567}$, corresponding to a $3.13\sigma$ level of confidence.

### 6.2.3 Omitting Weakest Flares

Figure 3(a) is a plot of the h values of the TS149 flares as defined in Section 5.1, ordered by the height of their peaks. Panel (b) is the same for the 120 G flare defined in Section 5.2. I now consider the set TS140, consisting of the *Spitzer* flares from which the 9 weakest ones, those that are to the left of the vertical line in panel (a), are removed. They are indicated by the % symbol in Table 2. I also consider now set TG115 of flares which is TG120 from which the 5 with the smallest h value are removed. They are the ones to the left of the vertical line in panel (b) and are indicated by the % symbol in Table 3. Set TIR255 is the combined set of TS140 and TG115.

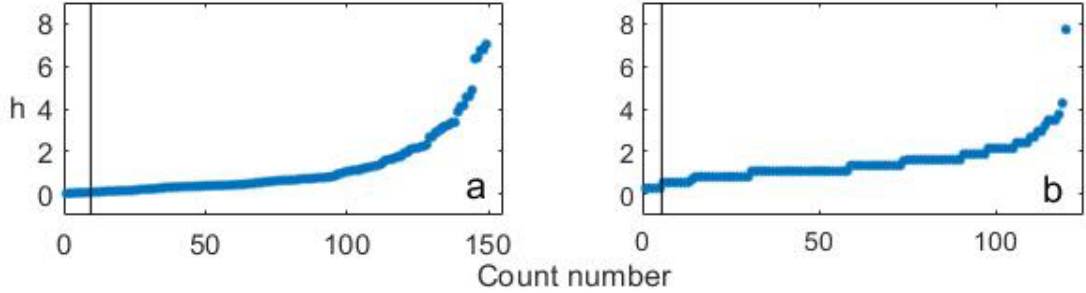

Figure 3: (a) Height h of peaks of 149 NIR flares recorded by *Spitzer*, in arbitrary, relative units. (b) Same as panel (a) for 120 NIR flares observed from the ground. Points to the left of the vertical lines are of the weakest flares referred to in the text.

The deepest minimum in the FDD of TIR255 is at f=35.3914 with the dispersion parameter value S=0.2411. The second-deepest minimum is at f=35.4005 with S=0.2466 and the third-deepest one is at f=35.3941 with S=0.2469. These numbers should be compared with the corresponding ones found for the TIR269 set.

Applying S-test on the TIR255 set with a sample of 322 SIR255 sets of PO events, we find $FNP_1 = \frac{12}{322}$. G-test of this file with the parameters 0.02<=w<=0.04 and K=250 of section 3.2.3 yields the estimate $FNP_2 \leq \frac{1}{322}$. The product of these 2 independent estimates is $FNP_{IR}^{(1)} = \frac{1}{8640}$, corresponding to a statistical confidence level of $3.86\sigma$.

We can also apply C-test on the TIR255 set. We consider its two independent subsets TS140 and TG115. The deepest minimum of TS140 is at $F_{140} = 35.1957$. S-test and G-test applied on this subset with a sample of 432 PO events yield the NFP estimates $\frac{1}{2.6}$ and $\frac{1}{9.2}$, respectively. The probability of falsely rejecting the null hypothesis regarding the TS140 set is the product of the two $NFP_{140} = \frac{1}{23.9}$.

The third-deepest minimum in the FDD of TS140, of the ordinal number $n_a = 3$, is at the frequency $f_a = 35.3967$. For subset TG115 a minimum at $f_b = 35.4005$ has the ordinal number $n_b = 10$ in the list of the deepestminima in its FDD. With expression (1) we find that the probability of the matching, within $\pm(f_b - f_a) = \pm 0.0038$ of the 2 frequencies, is $FNP(Ctest) = \frac{1}{210.4}$. Combined with the independent probability $FNP_{140}$ we obtain the estimate $FNP_{IR}^{(2)} \cong \frac{1}{5028}$ of falsely rejecting the null hypothesis regarding the pacemaker regulation of the TIR255 list of events. This probability corresponds to a $3.72\sigma$ level of confidence. Here again this result is not independent of the former estimate of confidence level of $3.86\sigma$ but it is not equivalent to it statistically.

A comment about the clipping of the extreme small flares from the TS149 and the TG120 data sets is warranted here. Maples et al. (2018) analyzed critically the common usage of "sigma clipping" or the Chauvenet criterion for rejecting outliers from a sample used to evaluate statistics of a population. The clipping procedure that I applied here on the 2 NIR flare lists does not fall within these categories. The methods described by Maples et al. use some statistical properties within the sample, quantified by the very same parameter the statistics of which in the population at large is being sought. Here the clipping process employs a parameter of the sample events, the height of the flares, which under the null hypothesis is independent of the property that we are investigating, namely, the distribution on the time axis of the flare times. The extreme nature of the removed flares from the 2 lists has most probably some physical origin but it is very unlikely that it has to do with the times of the flares, the statistics of which is the subject of our analysis. See further discussion of this point in Section 8.

## 7. Pacemaker vs. Periodicity Search

The flaring phenomenon of Sgr A*, as it appears in the time domain, has been analyzed extensively (e.g. Do et al. 2009; Meyer et al. 2009; Witzel et al. 2012, 2018 and references therein). Nearly all published investigations along this line have used various techniques of period search in time series.

The most commonly analyzed astronomical time series are LCs. They consist of a series of pairs of variables, an independent one, usually time, and a dependent one – magnitude or flux density. Here we are dealing with time series consisting of just one parameter, namely, time itself. We are looking for a statistical regularity in the distribution of numbers representing points along the time axis. Therefore the commonly used techniques in the analysis of LCs, such as power spectrum analysis or structure function (Simonetti et al. 1985; Hughes et al. 1992; Meyer et al. 2009; Emmanoulopoulos et al. 2010), are inadequate for performing the search that I am conducting. For this reason the regularities that I find as results of my search routine are not in conflict with statistical characteristics that are found in LCs with the usual techniques that seem to represent the same physical phenomenon, nor are they strengthening them.

In L2 I showed, with the help of a synthetic LC that I constructed around the times of 71 X-ray flares, that the structure of the power spectrum of the LC is entirely different from the structure of the FDD of the same basic phenomenon, computed on the same frequency search interval. Here too, I constructed a synthetic LC around the times of TG120 peak times as described in Section 6.2 in L2. The general structure of the PS shares basic common features, for example, with the structure function of IR LC presented in Meyer et al. (2009) but has no resemblance to the FDD of the TG120 time set. In particular, for example, the PS of the artificial LC of TG120 LC has no recognizable peak near $f=1$, the frequency which is the most outstanding feature in the FDD of that set of times. Similarly, the structure of the PS of the *Spitzer* LC, from which I extracted the TS149 time set, is entirely different from the general structure and in almost all details of the FDD of that time set.

## 8. Interpretation

The results of the analysis performed in the previous sections are products of a pure statistical inference applied on time series of measurements in a certain astronomical object. No reference is being made to the specific nature of this object. In particular our knowledge that the object is the BH at the center of Sgr A* played no role in this analysis. Therefore, to the level of the statistical confidence that we can attribute to the discovery of the regulation by the two pacemakers of the X-ray and the IR flares of Sgr A*, these phenomena require some interpretation, independent of the validity of the one that I suggest in this section.

The high statistical significance of the grouping of the X-ray flares of Sgr A* around the tick marks of the $F_X$ grid suggests that the pace making is maintained by some periodic physical process in the object astronomical system. A binary orbital motion is naturally suggesting itself, and in L1 and L2 I proposed that the pacemaker in the system is the periodic orbital motion of a mass around the central BH. In particular I suggested that the X-ray pacemaker period is the time interval between two successive passages of the mass through the pericenter point of the slightly eccentric precessing orbit of its revolution. This is frequently referred to as the epicyclic period ($P_{epi}$) of the system. (e.g. Kluzniak and Lee 2002; Abramowicz 2009).

The mass of the BH of the order of $\sim 4 \times 10^6\, M(Sun)$ (Gillessen et al. 2017) and the short period of ~149 min imply that the radius of the obit is of the order of a few Schwarzschild radii of the BH. I assume that the central object is a non or slowly rotating BH, embedded in spacetime of Schwarzschild metric. I shall comment on this assumption in the last section.

### 8.1. The Basic Post-Newtonian Equations

According to Abramowicz & Fragile (2013), stable Keplerian orbits do exist around non rotating BHs, of radii that are larger than the radius of the innermost stable circular orbit (ISCO). In the Schwarzschild metric it is equal to 3 Schwarzschild radii. Table 4 presents basic equations of motion of a small mass, call it a star, around a BH in a nearly Keplerian orbit of a small eccentricity e, expressed in a post Newtonian approximation. On the left hand side are the equations in which the gravitational potential at the orbit of the star, at distance r from the central BH of mass M, is given by the pure Newtonian expression $\varphi_N(r) = -\frac{GM}{r}$.

Paczynsky & Wiita (1980) suggested that in presenting the dynamics of a mass deep in the Schwarzschild metric, a better post Newtonian approximation to the general relativity (GR) equations can be achieved by replacing the pure Newtonian expression for the gravitation potential $\varphi_N = -\frac{GM}{r}$ with the potential $\varphi_{PW} = -\frac{GM}{r-R}$. Here $R = \frac{2GM}{c^2}$ is the Schwarzschild radius of the BH. This was found to be very useful by other researchers and its use is common in the literature (Abramowicz 200; Ruffert & Janka 201; Ohsuga & Mineshige 2011; Shakura & Lipunova 2018). The right hand side of Table 4 presents the post Newtonian equations where $\varphi_{PW}$ is assumed rather than $\varphi_N$.

In the table a parameter symbol X with double prime (X''), indicates the value of the parameter as measured in a frame revolving with the star around the BH. The symbol X' is the parameter value as measured by an observer at rest at a distance r from the center. The X symbol is the measured value by an observer on Earth.

Equations (2) in the table are the Kepler equations for a small mass revolving around a very large one, in an elliptical orbit with a semimajor axis a, and $x = \frac{a}{R}$. Here, $P_{orb}$ is the mean orbital period over many cycles (see Section 8.3.2). In Equations (3), e is the eccentricity of the orbit. They present the angle of the GR apsidal precession of the orbit, per one cycle of the revolution of the star, expressed as a fraction of the orbital cycle (Weinberg 1972). Equations (4) are the relation between the epicyclic and the orbital periods. Equations (5) are the epicyclic period as measured by an observer at rest at a distance r from the center, as affected by Special Relativity time dilation due to the high velocity v of the star in its orbital motion with respect to this observer. Here and in Equations (6), a is taken as the distance r of the star from the BH. A comment on this substitution will be made in Section 9. Equations (6) are the epicyclic period as measured on Earth due to the GR time dilation effect. Equations (7) are combining Equations (2), (3), (4) and (5) into Equation (6). Equations (8) are a rearrangement of Equations (7). Equations (9) are expressing Equations (8) as a dependence of M on x and $F_{epi}$. The numerical value in these equations is $987 = \frac{c^3}{4\sqrt{2}\pi G}\left(\frac{8.64 \times 10^4}{1.998 \times 10^{39}}\right)$. The expression in the brackets is a conversion factor for having M expressed in $10^6 M(sun)$ units and F in $\frac{1}{day}$ units. Equations (10) are rearrangements of equations (4). These equations are formally valid for x>2.8. With smaller x values M or $F_{orb}$ take negative values.

| | Newton | Paczynsky-Wiita |
|---|---|---|
| 2 | $P''^2_{orb} = \frac{4\pi^2}{GM}a^3 = \frac{4\pi^2 R^3}{GM}x^3$ | $P''^2_{orb} = \frac{4\pi^2(a-R)^2}{GM} = \frac{4\pi^2 R^3}{GM}x(x-1)^2$ |
| 3 | $\gamma = \frac{\delta\theta}{2\pi} = \left(\frac{1}{2\pi}\right)\frac{6\pi GM}{c^2 a(1-e^2)} = \frac{3}{2x(1-e^2)}$ | $\gamma = \left(\frac{1}{2\pi}\right)\frac{6\pi GM}{c^2(a-R)(1-e^2)} = \frac{3}{2(x-1)(1-e^2)}$ |
| 4 | $P''_{epi} = \frac{1}{1-\gamma}P''_{orb} = \frac{2x}{2x-\frac{3}{1-e^2}}P''_{orb}$ | $P''_{epi} = \frac{1}{1-\frac{3R}{2(a-R)(1-e^2)}}P''_{orb} = \frac{2(x-1)}{2(x-1)-\frac{3}{1-e^2}}P''_{orb}$ |
| 5 | $P'_{epi} = \frac{P''_{epy}}{\sqrt{1-\frac{v^2}{c^2}}} = \frac{P''_{epy}}{\sqrt{1-\frac{R}{2a}}} = \sqrt{\frac{2x}{2x-1}}P''_{epi}$ | $P'_{epi} = \frac{P''_{epy}}{\sqrt{1-\frac{Ra}{2(a-R)^2}}} = \sqrt{\frac{2(x-1)^2}{2(x-1)^2-x}}P''_{epi}$ |
| 6 | $P^2_{epi} = \left(\frac{1}{1-\frac{R}{a}}\right)P'^2_{epi} = \left(\frac{x}{x-1}\right)P'^2_{epi}$ | $P^2_{epi} = \left(\frac{1}{1-\frac{R}{a-R}}\right)P'^2_{epi} = \left(\frac{x-1}{x-2}\right)P'^2_{epi}$ |
| 7 | $P^2_{epi} = \left(\frac{x}{x-1}\right)\left(\frac{2x}{2x-1}\right)\left(\frac{2x}{2x-\frac{3}{1-e^2}}\right)^2 x^3\left(\frac{4\pi^2 R_S^3}{GM}\right)$ | $P^2_{epi} = \left(\frac{x-1}{x-2}\right)\left(\frac{2(x-1)^2}{2(x-1)^2-x}\right)\left(\frac{2(x-1)}{2(x-1)-\frac{3}{1-e^2}}\right)^2 x(x-1)^2\left(\frac{4\pi^3 R_S^3}{GM}\right)$ |
| 8 | $\frac{8x^7}{(x-1)(2x-1)\left[2x-\frac{3}{1-e^2}\right]^2} = \left(\frac{c^6}{32\pi^2 G^2}\right)\frac{P^2_{epi}}{M^2}$ | $\frac{8x(x-1)^7}{(x-2)(2x^2-5x+2)\left[2(x-1)-\frac{3}{1-e^2}\right]^2} = \left(\frac{c^6}{32\pi^2 G^2}\right)\frac{P^2_{epi}}{M^2}$ |
| 9 | $M = \left(2x - \frac{3}{1-e^2}\right)\sqrt{\frac{(x-1)(2x-1)}{8x^7}}\left(\frac{987}{F_{epi}}\right)$ | $M = \left[2(x-1) - \frac{3}{1-e^2}\right]\sqrt{\frac{(x-2)(2x^2-5x+2)}{8x(x-1)^7}}\left(\frac{987}{F_{epi}}\right)$ |
| 10 | $F_{orb} = \frac{2x}{2x-\frac{3}{1-e^2}}F_{epi}$ | $F_{orb} = \frac{2(x-1)}{2(x-1)-\frac{3}{1-e^2}}F_{epi}$ |

Table 4: Post Newtonian expressions, approximating the GR equations that describe the dynamics of a small mass near the BH. On the left-hand side the gravitation potential is approximated by the Newtonian function $\varphi_N = -\frac{GM}{r}$. On the right-hand side the potential function is the Paczynsky-Witta one $\varphi_{PW} = -\frac{GM}{r-R}$. See text for explanations.

## 8.2. The Parameters of the Star Motion

According to the interpretation suggested here, $F_X = 9.8697$ of Section 4.1 is identified as $F_{edi}$. The frequency $F_{IR} = 35.395$ of Section 6.1 is identified as $F_{orb}$. The midpoints of the X-ray flares of set TX81 seem to be the most secure data set since nearly no personal bias or subjective selection were involved in their determination. Also the $F_X$ frequency seems to be highly significant statistically. Therefore, taking $F_{edi} = 9.6897$ as given, we may consider expressions (9) and (10) in the table as equations for M and for $F_{orb}$ as functions of x. From their combination we can derive a direct dependence of $F_{orb}$ on M.

Figure 4 presents the graphs of these functions as obtained from the equations on the right-hand side of Table 4. Panel (a) depicts the mass M of the BH as a function of the orbital semimajor x, to which we shall refer as the radius of the orbit. Panel (b) is the orbital frequency vs. the orbital radius. Panel (c) is the BH mass vs. the orbital frequency. The black curves are for $e \approx 0$ and the red ones are for e=[0.1, 0.2, 0.3, 0.35, 0.4]. The 2 horizontal lines in panel (a) and the 2 vertical lines in panel (c), at M(min)=3.9 and M(max)=4.4, delimit the range within which the mass of the BH at the center of Sgr A* has been estimated by various researchers in the last few years (Ghez et al. 2008; Genzel et al. 2010; Gillesen et al. 2006, 2017; Boehle et al. 2016; Parsa et al. 2017). The two horizontal thin lines in panel (c) at $F_{orb}(min) = 24$ and $F_{orb}(max) = 48$ are the limits of the uncertainty interval in the value of the orbital frequency of the moving mass that was detected by Abuter et al. (2020 – GRAV) as estimated by these authors.

The thick horizontal blue line marks the $F_{IR} = 35.395$ value. It is the frequency of the NIR pacemaker as extracted from the measured NIR flares timing, as described in Section 6.1. Within the context of the model this number is identified as $F_{orb}$ and panel (c) shows that it falls almost at the center of the range of possible $F_{orb}$ values as estimated from the direct observations of GRAV. The $F_{IR}$ value corresponds to an orbital period $P_{orb} = 40.7\,min$ while GRAV suggest $P_{orb} = 45\,min$. At $F_{orb} \cong 35.4$ the orbital velocity of the star is $v \approx 0.3c$.

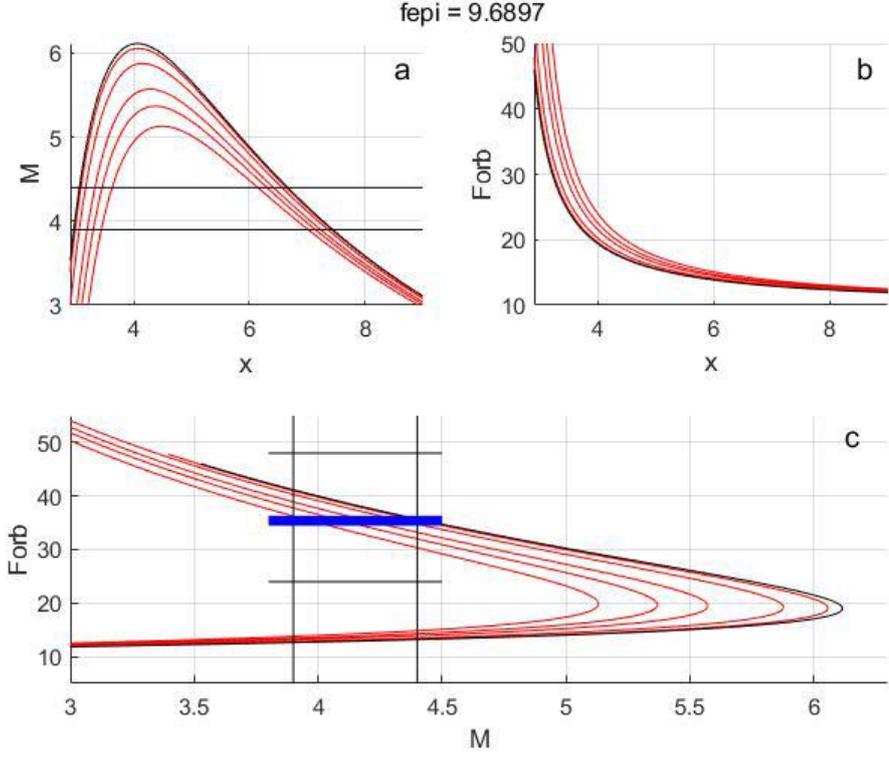

Figure 4: (a) The functional relations (a) between the BH mass M($[10^6 M(Sun)]$) and the orbital radius $x = \frac{a}{R}$, (b) between the orbital frequency $F_{orb}$ and x and (c) between $F_{orb}$ and M. They represent equations (9) and (10) that are on the right-hand side of Table 4, taking $F_{epi} = 9.6897$ as given. The black curve is for eccentricity of the orbit $e \approx 0$. The red curves are for e=[0.1, 0.2, 0.3, 0.35, 0.4]. The thick horizontal blue line marks the value $F_{IR} = 35.395$ as derived from the IR flares timing by the FDD analysis. The 2 horizontal lines in panel (a) and the two vertical lines in panel (c) delimit the region of BH mass determinations by various observers during past few years. The horizontal thin lines in panel (c) delimit the uncertainty in F_orb as stated by GRAV. See text for further explanations.

Table 5 presents the M coordinates of the intersection points of the curves in panel © with the $F_{IR}$ blue line in the figure. It also presents the x values corresponding to these M values.

| e | 0 | 0.1 | 0.2 | 0.3 | 0.35 | 0.4 |
|---|---|-----|-----|-----|------|-----|
| M | 4.42 | 4.41 | 4.34 | 4.2 | 4.1 | 3.97 |
| x | 3.07 | 3.08 | 3.15 | 3.27 | 3.36 | 3.47 |

Table 5: Mass M of the BH, in $10^6$ M(Sun) Units, and the semimajor axis $x = a/R$ of the orbit of the star revolving around it, for different eccentricities e

of the orbit, as implied by the model presented in this work. Here the equations on the right-hand side of Table 4 are employed. These parameter values are obtained assuming that the epicyclic frequency is $F_{epi}=F_X=9.6897$ and the orbital frequency is $F_{orb}=F_{IR}=35.395$.

The table shows that if we took at face value the pacemakers frequencies $F_X$ and $F_{IR}$ as derived by the analysis presented in this work, and if we adopted the proposed model by identifying $F_X = F_{epi}$ and $F_{IR} = F_{orb}$, assuming $0 \leq e \leq 0.4$, we would predict that the mass of the BH falls precisely within the interval of the uncertainty in its value set in the last few years by various observers of the Sgr A* environment. But see a critical comment on this table in Section 9.

The numbers in Table 5 remain nearly unchanged when we take for $F_{epi}$ the values 9.66 or 9.72 which delimit the uncertainty interval in the value of $F_X$, as explained in Section 4.1.

It should also be noted that when the model is represented by the equations on the left-hand side of Table 4, i.e. when in the post Newtonian approximation for the GR equations, the pure Newtonian potential is utilized rather than the Paczynsky-Witta one, the model is not compatible with other observational data. In that case, assuming, as we did, that $F_X = F_{epi} = 9.6897$ and $F_{IR} = F_{orb} \approx 35.4$, Equations (9) and (10) on the left-hand side do not give pairs of M and x values that are consistent with uncertainty limits on the values of these parameters that are well established by observations.

Panel (a) of Figure 4 may explain why in L1 and L2 an orbital radius $x \approx 6.6$ was suggested. The figure shows that for the range of the uncertainty in the value of the BH mass, delimited by the two horizontal lines in panel (a), for the same $F_{epi}$ value there are two distinct regions of possible x values, one around x=3 and one around x=6.6. The finding in this work of the pacemaker signal of $F_{IR} \approx 35.4$ in the NIR flares of the object, combined with the $F_{orb}/x$ relation presented in panel (b), removes this degeneracy. Lu et al. (2018) report on the detection by very large baseline interferometric observations of a compact intrinsic radio source structure on scales of ~3 Schwarzschild radii of the BH. This finding is very much in line with the x values in Table 5 although the authors interpret the structure in the context of brightness distribution of disk and jet-dominated model. And there is now the direct measurement by GRAV of the 45 min periodicity of the mass revolving around the BH with its associated $x \approx 3$ value.

The possibility, raised in L2, that the revolving mass is a planet, is also becoming now implausible. With expression (2) presented by Eggleton (1983) one can calculate the Roche lobe radius of a mass at a given distance from the $4 \times 10^6 M(Sun)$ BH of Sgr A*. From the mass-radius relations obeyed by hundreds of exoplanets (Bashi et al. 2017; Kanodia et al. 2019; Otegi et al. 2020) it appears that for all planetary masses, at the distance $x \sim 3$, the planet radius is much larger than the Roche lobe radius. Therefore planets, as we recognize these objects, have no stable existence in that environment.

## 8.3. The Mechanisms of the Pacemakers

### 8.3.1. The X-Ray Pacemaker

We know that there are quite a few stars of the S cluster that revolve around the BH at the center of Sgr A*. Yusef-Zadeh et al. (2017) presented evidence for low-mass star formation in close vicinity to the BH. A cusp of Late-type Stars around the central BH has been investigated also by Habibi et al. (2019). The objects that these authors have identified lie 2 orders of magnitude further away from the BH than the few gravitational radii of the orbit of the mass discussed here, but it seems that it is the angular resolution limits of the observations that set the scale of distances explored by these authors. The revolving mass could be a member of this class of newly formed low mass stars. In fact, it could be one component of an originally binary system that was part of the cusp of the low mass stars around the center. The binary system underwent an exchange collision with the BH, leaving one component revolving around the BH while the other was ejected outward to become a high velocity star in the Galaxy (Hills 1988, Koposov et al. 2019).

According to the model suggested here, the X-ray flares of Sgr A* are episodes of intense mass loss from the star. They are triggered by large deformations of the outer layers of the star by the strong tidal forces acting on it. The resulting violent tidal waves are driven at the epicyclic frequency of the binary system. Every 9 or 10 cycles of this frequency, on the average, the tidal waves in the outer layers of the star reach such large amplitudes that when the star is close to the phase of pericenter passage, when its Roche lobe radius takes its minimum value, mass loss becomes especially intense, giving rise to the observed X-ray flares. The difference t(end)-t(begin) of an X-ray flare is the duration time of the mass loss episode and the midpoint between these 2 times is an appropriate marker of the time of the event. The flare peak time is not an adequate marker since the instantaneous mass loss rate during a flare depends strongly on the wave fronts of the violent tidal waves in the outer layers of the star at that time. King (2020) has recently suggested the very same mechanism as an explanation of the 9 hour X-ray quasi-periodic eruptions (QPEs) from the nucleus of the Seyfert 2 galaxy GSN 069 detected by Miniutti et al. (2019).This mechanism may also be the explanation of the recently discoveredQPEs from the galactic nucleus ofRX J1301.9+2747 (Giustini et al. 2020).

### 8.3.2. The NIR pacemaker

While the $F_X$ pacemaker affecting the X-ray flares is a modulation of the major energy source of these flares at the epicyclic frequency, with the IR pacemaker the situation is different. The origin of the IR flares, along with those in X-ray, is a much discussed theme in the literature (Eckart et al. 2004; Dodds-Eden et al. 2009, 2011; Yusef-Zadeh et al. 2009; Younsi and Wu 2015; Li et al. 2017; Chael et al. 2018; Eckart et al. 2018; Witzel et al. 2018; Ripperda et al. 2020 and references therein). However, as stated by many researchers in the field the cause of the erratic outbursts of the flares or the physical origin of the episodic acceleration of electrons that presumably gives rise to the flare outbursts are not well understood (e.g. Barriere et al. 2014, Zhang et al. 2017, Ripperda et al. 2020).

Based on detailed measurements of flares that were conducted simultaneously in the X-ray and the NIR spectral regions it was suggested that flares, at least large ones, start first in the NIR radiation while X-rays are detected only a few, or a few tens of seconds later (Yusef-Zadeh et al. 2006,2012, Ponti el al. 2017). Recently, Boyce et al. (2019) also found in a sample of simultaneous multi-wavelength measurements that there is a rise in IR flux density around the same time as that of every distinct X-ray flare. However, in contrast to previous suggestions these authors announced a detection of a possible lag of a few minutes of IR flares after the onset of the X-ray rise. There is not necessarily a real conflict between these apparent conflicting reports as all the measured time differences between IR and X-ray flares seem to be within the uncertainty in the measurements of these differences. In particular is was reported by the observers that with these uncertainties, all differences are consistent with no measureable difference.

Boyce et al. state, that the correlation between the times of IR and X-ray flare is in one direction only. Peaks in the IR emission may not be coincident with an X-ray flare. An X-ray flare might be a sufficient condition for an outburst of a NIR one, although this has not yet been established observationally well enough, but it is clearly not a necessary one. It is enough to inspect the *Spitzer* LC parts of which are accompanied by X-ray observations and to realize that a number of IR flares, including large ones, have no X-ray predecessor along tens of minutes preceding them . Also in the still rather small sample of simultaneous observations, there seems to be no obvious correlation between the amplitude, the duration or the fluence of the NIR flares with those of the X-ray ones.

It may therefore be not implausible to suggest that the initial trigger of a NIR flare is not directly connected to the X-ray production mechanism.

Following previous suggestions (Broderick & Loeb 2006; Eckart et al. 2008a; Li et al. 2017; Ponti et al. 2017) GRAV proposed that magnetic shocks and reconnections accelerate electrons to highly relativistic energies that generate the IR flares. GRAV and others suggested an analogy with the processes at the origin of solar flares (Dodds-Eden et al. 2010; Ponti et al. 2017). What is proposed in the model presented here is that this analogy may be taken one step further, proposing that these processes are indeed taking place within the outer layers of the revolving star, not unlike the known flaring activity of stars of various spectral and luminosity types (e.g. Schaefer et al. 2000, Davenport et al. 2014, Chang et al. 2020). In this context it could also be suggested that the origin of the recent unusual variability of the NIR luminosity of Sgr A* reported by Do et al. (2019) may beakin to the megaflare phenomenon observed in stars  (Anfinogentov et al.2013). The main energy source of the Sgr A* NIR flares could also be in hot spots in the accretion disk around the center that are at close vicinity of the star and are excited by some process that takes place in the outer layers of the star. The main point here is the suggestion that the principal timing of the outbursts, as well as their amplitudes and duration, are determined by physical processes that are internal to the star and are not directly related to the frequency of its binary orbit. GRAV state that "all three flares [the data of which are analyzed in their paper] can in principle be accounted for by a single orbit model". Tracing the roots of the flares in internal processes within the star explains well the physical origin of the spots as well as their confinement to a single orbit around the center.

A one-way correlation between occurrence times of X-ray flares and NIR flares that are contemporaneous or immediately following them may well be consistent with this proposition. The dynamo processes that are believed to be at the roots of the stellar flaring phenomenon are intimately connected with the magnetohydronamics of the star layers where flares are initiated (see for example a review by Benz and Gudel 2010). An X-ray flare is generated when a specially intense mass-loss event takes place, mainly through the L1 point of the Roche lobe of the star. At this time the tidal driven dynamics of the outer layers of the star is changing its character from a circularly closed or quasi standing waves pattern to an open end streaming flow. This may be an extra stimulus for initiating the process that gives rise to a NIR flare.

Superposed on the erratic flaring variability originated in the star there is some weak modulation that operates not on the energy source of the flare but on the radiation transfer of its emitted radiation. There is a cyclic variation with the orbital frequency in the aspect ratio of the star and the BH with respect to the line of sight from Earth as the star revolves around the BH. Beaming, lensing and time dilation in the vicinity of the BH, affect the flare radiation in the direction of Earth as the light beam passes through this cyclically varying environment.

Figure 1 in Hamaus et al. (2009) presents a computed LC of a compact IR radiation source in a circular motion around the BH of Sgr A*. For an orbit with an inclination $i = 20^o$ the amplitude of the periodic variation in the apparent magnitude of the radiation source reaches 1 magnitude. This means that for an inclined orbit of the object there is a section of the orbital cycle during which the probability of detecting a flare is smaller than in the other section of the cycle.

The nature of the pacemaker with the $F_{IR}$ frequency is then the selection effect in the number of the detected flare events, introduced by these varying radiation transfer conditions. As opposed to the X-ray flares, here the relevant parameter that determines the detectability of a flare is the flare peak luminosity. The very weak flares that erupt while the star is in the orbital phase section of low luminosity as seen from Earth, are missed by Earthlings. This also explains why removing the very weak flares from sets TS149 and TG120, as described in Section 6.2.3, has strengthened the $F_{IR}$ pacemaker signal in the data. Some of the 14 apparent weak flares that were removed are likely to be intrinsically slightly brighter ones that had erupted at the low luminosity phase section of the star orbit. By removing these flares we have increased the imbalance between the number of detected flares emitted during the "dark" phase section of the orbit and the number emitted during its "bright" section. This has accentuated the pacemaker signal. The relative weakness of this selection effect, manifested by the quite large dispersion of the NIR events around tick marks of the orbital frequency grid, indicates that the orbit of the star is seen nearly pole on. This is very much in line with the conclusion arrived at by GRAV from their proper motion measurements.

## 9. Critical and concluding remarks

The main claim made in this paper is the uncovering of statistical regularities in the measured occurrence times of the X-ray and the NIR flares of Sgr A*. These regularities were referred to as modulation by the $P_X$ and the $P_{IR}$ pacemakers. This claim stands on its own, regardless of the validity of the model suggested as an interpretation of the phenomenon. It rises or falls on the merits of the statistical evidence that can be mastered in support of it. This paper shows that the claim can be established with confidence level of more than $4.6\sigma$ for the X-ray flares and at more than $3.8\sigma$ for the IR ones. Nonetheless, one cannot rule out completely the possibility that one or the 2 pacemaker effects found in the data are statistical flukes after all.

A weak link in the major claim is the possible systematic yet unknown bias in the selection of the most basic data sets, namely, the HJD dates of the midpoints of the X-ray flares and of the peak points of the IR flares that were taken as the events to be analyzed. This is why additional observations and analysis by different people are required in order to strengthen the evidence for the operation of the two pacemakers, or to refute this claim.

The numbers $S_X = 0.1976$ (Sections 4.1) and $S_{IR} = 0.2411$ (Section 6.2.3), should be compared with 0.2887, the StD of a rectangular distribution over the [-0.5,0.5] interval. It seemsthat the grouping of the X-ray flares, and even more so of the NIR ones, around tick marks of the corresponding pacemakers are not very tight. The model presented in Section 8 suggests that the dispersion of the flare times around the pacemakers tick marks is inherent to the flare phenomenon itself. Rather than being well concentrated around some fixed phase of the pacemaker cycle the flare events occur within a wide section of it. Therefore, if future observations are to strengthen the evidence for pacemakers in the Sgr A* system it will not come in the form of diminishing the value of the S parameter as longer series of detected flare times are considered. Further evidence could come from finding in future new independent sets of observations the effect of two pacemakers of the same frequencies $F_X$ and $F_{IR}$, but not necessarily with smaller dispersions.

On the theoretical side, the model suggested here relies on post-Newtonian equations applied on a highly relativistic system. Such an approach could be too simplistic. For one thing, there is, so far, no observational hint regarding the rate of rotation of the BH at the center. If it is a fast rotator, the Schwarzschild metric itself is not an appropriate description of spacetime in its vicinity. But even within the context of the Schwarzschild metric it is yet to be proven that the Paczynsky-Witta potential function is an adequate approximation in the extreme conditions of existence of the alleged star.

Furthermore, the equations in Table 4 are strictly consistent only for $e \approx 0$ implying, according to Table 5, that $M \cong 4.4$. This value lies at the extreme end of the uncertainty in the value of M as estimated by other, more direct measurements. Also, as noted in Section 8.3.2, $P_{orb}$ must be understood as referring to a mean value of the sidereal orbital cycle of the star, and the NIR pacemaker frequency $F_{IR}$ uncovered in the data is the mark of this mean value. In Equations (5) and (6), the relevant parameter x should represent an instantaneous distance r of the star from the BH that is varying between the extreme values $a(1+e)$ and $a(1-e)$ at the apicyclic frequency. For $e > 0$, the use of the constant a value in these equations may therefore be a poor approximation. Also for $e > 0.02$ the orbit leads the star to within the ISCO circle around the BH where the post-Newtonian approximation may break down. It may therefore be that in order to place the model on firmer grounds, a full relativistic treatment, or the use of equations that prove to be adequate for describing the system as assumed in the model, may be required.

Most recently, Dexter et al. (2020), presented very detailed models of various magneto-hydrodynaminal processes within a framework of mass accretion disk at a few gravitational radii from Sgr A* BH. The models provide an explanation of various observed characteristics of the object radiation, from radio to NIR frequencies. These include also the highly variable near-infrared flaring emission, which is, along with the X-ray variability, the subject matter of this work. It seems, however, that the models do not provide an explanation of the main results of this work, namely the statistical, pacemaker regularities in the flaring activity of the object in these 2 regions of the electromagnetic spectrum. The model suggested here is not in conflict with these and other physical models attempting to interpret measured properties of Sgr A* radiation. These models of micro physical calculations may perhaps be incorporated into the macro astronomical scenario suggested in this work. An example is the work by Yuan et al. (2003) who proposed that the non thermal X-ray emission of Sgr A*, particularly in its flaring mode, is due to occasionally enhanced particle acceleration in the inner region of the accretion flow around the BH at $r < 10R$. The required occasional enhancements may well be the short events of mass loss from the semi-detached binary companion that revolves around the BH, as suggested in the second part of this paper.

Finally, it may be hoped that in the least, this paper will encourage observers of Sgr A* to measure and publish exact times of occurrence of X-ray andNIR flares of this most interesting celestial object.